\documentclass[%
 aip,
 amsmath,amssymb,
 reprint,%
]{revtex4-1}

\usepackage[utf8]{inputenc}
\usepackage[T1]{fontenc}
\usepackage{mathptmx}

\usepackage{graphicx}
\graphicspath{{./figures/}}

\usepackage{dcolumn}
\usepackage{epstopdf}
\usepackage{bm}
\usepackage{amsmath}
\usepackage{mhchem}
\usepackage{xcolor}

\usepackage{color}

\newcommand{\PZT}{Pb(Zr$_{x}$Ti$_{1-x}$)O$_3$}

\newcommand{\PZTth}{PbZr${_{0.7}}$Ti${_{0.3}}$O$_3$}

\newcommand{\PZTfifteen}{PbZr${_{0.85}}$Ti${_{0.15}}$O$_3$}
\newcommand{\PZTten}{PbZr${_{0.9}}$Ti${_{0.1}}$O$_3$}
\newcommand{\PZTfive}{PbZr${_{0.95}}$Ti${_{0.05}}$O$_3$}

\begin{document}                  
\preprint{AIP/123-QED}

\title{Giant electrocaloric effect at the antiferroelecrtric-to-ferroelectric  phase boundary in \PZT}

\author{A. V. Kimmel}
\thanks{\textit{Corresponding author} a.kimmel@nanogune.eu}
\affiliation{CIC nanoGUNE, Tolosa Hiribidea, 76, San Sebastian, 20018, Spain}
\affiliation{Department of Physics and Astronomy, University College London, Gower Street, London WC1E 6BT, UK}
\author{O. T. Gindele}
\affiliation{Department of Physics and Astronomy, University College London, Gower Street, London WC1E 6BT, UK}
\author{D. M. Duffy}
\affiliation{Department of Physics and Astronomy, University College London, Gower Street, London WC1E 6BT, UK}
\author{R.E. Cohen}
\affiliation{Extreme Materials Initiative, Geophysical Laboratory, Carnegie Institution for Science, Washington, DC 20015, USA}
\affiliation{Department of Physics and Astronomy, University College London, Gower Street, London WC1E 6BT, UK}
\affiliation{Department of Earth and Environmental Sciences, Ludwig-Maximilians Universit$\ddot{a}$t M$\ddot{u}$nchen, Theresienstr,
41 80333 Munich, Germany}

\begin{abstract}
Molecular dynamics simulations predict a giant electrocaloric effect at the ferroelectric-antiferroelectric phase boundary in PZT (PbTiO$_3$-PbZrO$_3$).  These large-scale simulations also give insights into the atomistic mechanisms of the electrocaloric effect in \PZT.  
%Studying a range of PZT compositions we found positive electrocaloric effect in ferroelectric compositions,  but antiferroelecric ones exhibit a positive-to-negative crossover.
We predict a positive electrocaloric effect in ferroelectric PZT, but antiferroelectric PZT exhibits a negative to positive crossover with increasing temperature or electric field.
At the antiferroelectric-to-ferroelectric phase boundary we find complex domain patterns. 
We demonstrate that  the origin of giant electrocaloric change of temperature is related to the easy structural response of the dipolar system to external stimuli in the transition region.
\end{abstract}

\maketitle                      

%\section{Introduction}
\indent The electrocaloric 
effect is a reversible temperature change  ($\Delta T$) in  materials under adiabatic 
conditions in response to applied electric (or magnetic) field. 
The discovery of a giant 12 K  electrocaloric effect (ECE) in  thin films of Zr-rich lead titanate 
compositions  fuelled interest into the development of novel ferroelectric-based 
ECE materials\cite{Mischenko2006}.  

Giant and moderate ECE's have since been reported 
for classical ferroelectrics like BaTiO$_3$\cite{Kar-Narayan2010} and for several relaxor 
materials \cite{Lu2010}.
 Pb(Zr${_{1-x}}$Ti${_x}$)O${_3}$ (PZT) is a disordered solid solution ABO$_3$ perovskite, 
 with Pb atoms occupying the A-site, and Ti and Zr cations randomly arranged among the B-sites.  
PbTiO$_3$ (PTO), the  $x$=0.0 end member of \PZT,  is a classical ferroelectric (FE), and the other end member PbZrO$_3$ (PZO) ($x$=1.0) is antiferroelectric (AFE).  Near $x$=0.95 there is a  phase boundary that separates AFE and 
FE phases\cite{Woodward2005}. 
 Pb(Zr${_{1-x}}$Ti${_x}$)O${_3}$ (PZT) remains an active area of research for novel ECE 
 materials\cite{Zhang2016, Zuo2015}. 
The response of PZT to the applied electric field  in the  transition  region between its ferroelectric  
and antiferroelectric phases  is of particular interest since a giant electrocaloric response has 
been found experimentally for compositions close to this region\cite{Mischenko2006}. 
%%%%%%%%%%
%GENG et al
Studies of electrocaloric response of AFE Pb$_{0.97}$La$_{0.02}($Zr$_{0.95}$Ti$_{0.05}$)O${_3}$ 
have  provided an insight into a mechanism for the negative electrocaloric response. Authors  suggested  that
misaligning of non-collinear dipoles  provides different entropy contribution 
depending on the direction of the applied electric field  ~\cite{Geng2015}.
%%%%%%%%%%

Several theoretical works discuss  caloric effects in perovskites.
Large electrocaloric effects have been  observed in the vicinity of ferroelectric-paraelectric phase 
transition, however, little is known about the ECE near AFE-FE phase boundary.
Recent work with effective Hamiltonians reveals a strong potential of electrocalorics for thin PZO 
films with FE and AFE phase competition\cite{Ponomareva2018}. 
Phenomenological modelling  for an AFE system predicted the 
negative electrocaloric effect in PZO ceramics, which  agrees well with direct measurements 
of the EC temperature change in this system\cite{Pirc2014}.

Molecular dynamics (MD) methods, using shell model potentials fit to first principles calculations, 
are promising models for computing the thermal behaviour of materials,  since they do 
not require assumptions about the important degrees of freedom. 
Such models have been  used  to study ECE 
in LiNbO$_3$, PMN-PT, and BaTiO${_3}$\cite{Rose2012, *Erratum14, WuCohen2017a, WuCohen2017b}. 
These simulations provide insight into the universal principles related to optimal operating temperature for the electrocaloric effect. 

%%%%%%%
In this work we studied the effects of composition on electrocaloric properties of PZT 
using large scale MD simulations with first-principles based shell model  potentials\cite{Gindele2015}. 
We modelled a wide range of ferroelectric and antiferroelectric  compositions of \PZT. 
We found  that the electrocaloric response of  PZT  correlates with the type of ferroelectric 
order and that a giant electrocaloric response exists at the phase boundary of PZT, 
where antiferroelectric and ferroelectric order coexist.

%\section{Methods}
To model the electrocaloric properties of PZT   we use a 
core-shell  force field, which includes all degrees of freedom. 
This $ab$~ $initio$ based interatomic 
potential reproduces a set of temperature and composition induced phase transitions 
characteristic of \PZT \cite{Gindele2015}. 
The potential model underestimates the Curie temperatures with respect 
to experiment for PbTiO$_3$ (600 K versus 750 K\cite{Woodward2005}) and 
PbZrO$_3$ (400 K versus ~507 K\cite{Pirc2014}), which is a
reasonable error for this type of force field. 

A set of \PZT~compositions  were modelled using the DL$\textunderscore$POLY  code\cite{Todorov2006}. 
We study AFE and FE  compositions with  $x$ equal to 0.5, 0.9,  1 (corresponds to AFE PbZrO$_3$), together with  
 $x$=0 (that corresponds to FE PbTiO$_3$), 0.7, 0.8, 0.85, 0.95 shown in Supplementary Information (SI). 
The B-site cations, Ti and Zr,  were  randomly distributed over the B-sites.
We use the adiabatic shell model (also known as dynamical model~\cite{Fincham93}) as a method 
of incorporating polarisability into a molecular dynamics simulation 
with the shell masses  varying as 3.5 \%, 8.3 \%, 17.12 \% and 12.5 \% of the atomic mass of Pb, Ti, Zr and O, respectively.
We used relatively large 20 $\times$ 20 $\times$ 20 super-cells (80 000 core and 
shell particles).  Each composition was equilibrated at 100 K for 40 ps, followed by application of  
an electric field along the polar axis.  The direction of the polar axis depends on 
the composition of \PZT~ and was taken as [001] for PZO, [111] for the Zr content 
from 0.95 to 0.50, and as [001] for $x>=0.4$ . 
The strength of the applied electric field was 0, 5, 10, 15, 20, 25, 50, 75, 100 MV/m. 
We used a 0.2 fs timestep and 
NST ensemble with the Nos\'{e}-Hoover thermostat (0.01 ps) and barostat (0.5 ps) 
for equilibration of individual systems during 8 ps. The equilibration was followed 
by a 12 ps production run over which the polarisation value was calculated.  

To study the electrocaloric effect we used the indirect method, where the 
change of temperatures were calculated from Maxwell related expression:
\begin{equation}\label{eq:ECE2}
\Delta T=- \int \limits_{0}^{E} \frac{TV}{C_{p,E}} \left(\frac{\partial P}{\partial T}\right)_E  \mathrm{d}E,
\end{equation}
Here, $E$ is the applied electric field, $T$ is the temperature, $V$ is the volume of the 
simulation cell and $C_{p,E}$ is the heat capacity per cell under constant electric field and pressure. 
We calculate the ECE change of temperature ($\Delta T$),  by integrating  equation (\ref{eq:ECE2}) numerically.
The values of $C_{p,E}$  were calculated as the derivative of the total energy with temperature ($\frac{\partial E_{tot}}{\partial T}$) 
at a given value of electric field, $E$ and are in agreement with experiment ~\cite{Morimoto2003}  (See Supplementary Information).

%\section{Results}
%%%%%%%%%%%%%%%%%%%%
%%% PTO
%%%%%%%%%%%%%%%%%%%%
The temperature and field dependence of the electrocaloric change in temperature, $\Delta T$, were calculated for FE PTO  via expression  eq.~(1) ( see Supplementary Information Fig. 1a.) 
A characteristic dominant peak at 650 K  (the PTO Curie temperature reproduced by our force field) moves towards higher temperatures for larger applied electric fields, typical for ferroelectrics \cite{Rose2012,*Erratum14}. 
The magnitude of electrocaloric effect calculated for PTO  is good agreement with similar method computations for LiNbO$_3$  that gives 17 K at applied 50 MV/m field versus 16 K in our computations for PTO\cite{Rose2012}.

%%%%%%%%%%%%%%%%%%%%
%%%%   PZT50
%%%%%%%%%%%%%%%%%%%%
The morphotropic phase boundary (MPB) is found in a narrow compositional range around $x$=0.5,  where the FE phase with  rhombohedral symmetry  transforms to the tetragonal phase. 
It is now known that there is a monoclinic transition region between the rhombohedral and tetragonal phases ~\cite{Noheda1999, *Cohen2018, *Glazer2014, *Bogdanov2016, *FuCohen2000, *Ahart2008}.
We found that  the electrocaloric effect in FE  \PZT~  with  $x$ = 0.5 and 0.7 exhibits very similar behaviour.
The peaks of $\Delta T$ broaden, which reflects the B-site cations disorder and reduction of the correlation length in the material \cite{Rica2016} (See Fig. 1a, b and Supplementary Information, Fig. 1b, c). 
The $\Delta T$ curves peak above T$_c$ with increasing electric field (Fig.~1b), similar to what was computed for  LiNbO$_3$\cite{Rose2012}.\ 

%%%%%% FIGURE AFE dT %%%%%%%%%%%
  \begin{figure}[h]
  \centering
   \includegraphics[clip=true,width=0.45 \textwidth]{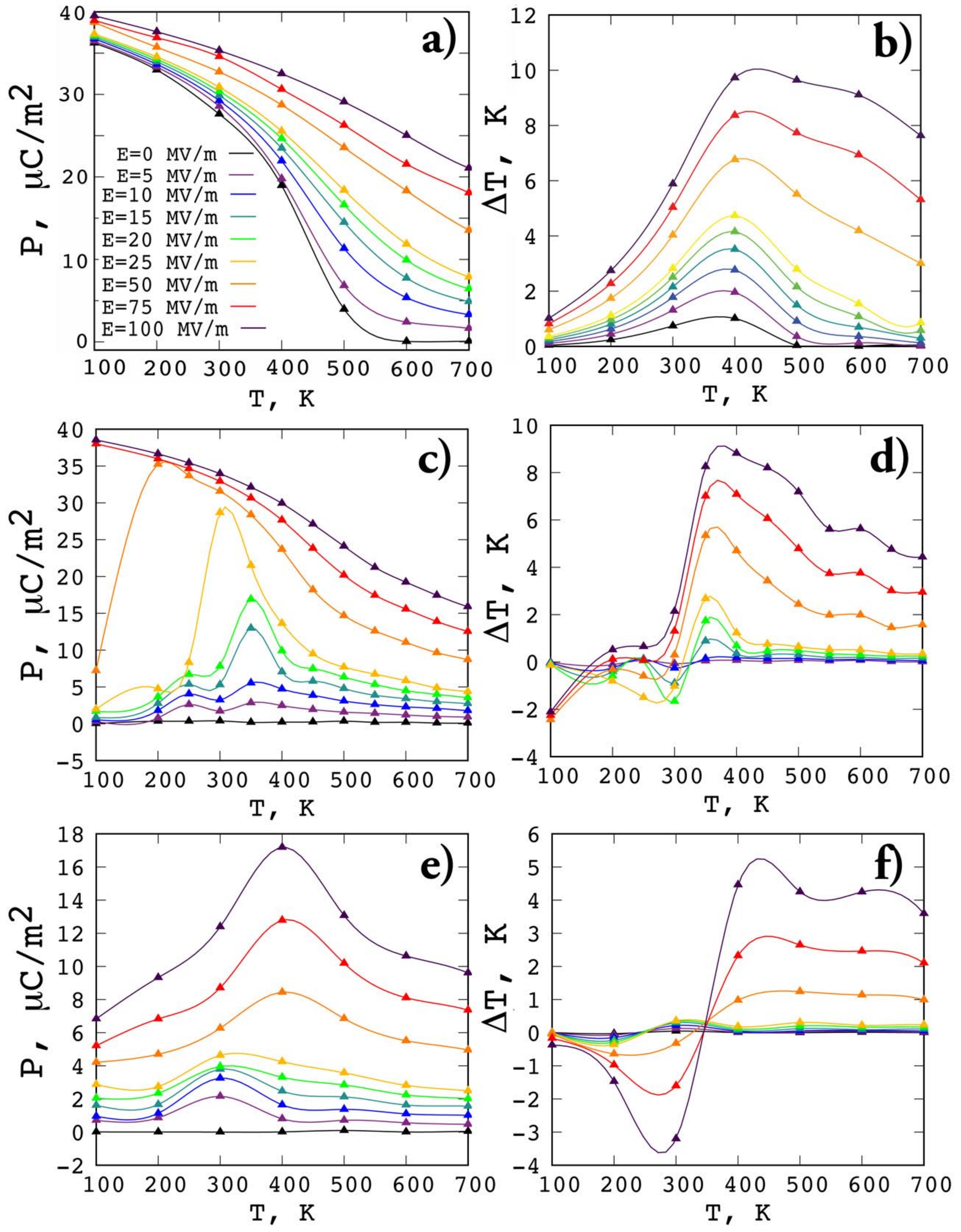}
    \caption{Polarisation and EC temperature change with temperature and different applied electric fields for  
    a), b) PbZr$_{0.5}$Ti$_{0.5}$O$_3$, c), d) PbZr$_{0.9}$Ti$_{0.1}$O$_3$ and e), f) PbZrO$_3$.
   }
    \label{fig:PZT_5-10-15-P}
  \end{figure}  

%%%%%%%%%%%%%%%%%%%%
%%%%%%% AFE
%%%%%%%%%%%%%%%%%%%%
The transition boundary between AFE and FE  phases  in \PZT~ has been shown 
to exist within a composition region around $x$=0.95-0.9\cite{Woodward2005}.  
It is challenging to  identify the precise composition of the transition
 region between  AFE and FE  phases experimentally, due to purity of the samples, 
 composition variance, especially for solid solution materials, and the presence of 
 surface effects that may stabilise the FE phase.
The force field used in this work is able to  reproduce the composition induced 
AFE-FE phase boundary, but the model gives a boundary wider than seen experimentally
 -- we find  composites with $x>$0.8 exhibit  antiferroelectric properties\cite{Gindele2015}.  
Further, we have performed calculations of the electrocaloric properties for several of the 
AFE PZT compositions  with $x$ of  0.9, and 1 (PbZrO$_3$), 
while the result for 0.85, 0.95 are given in Supplementary Information. 

We found that the electrocaloric response of AFE's  is very different 
to that of the FE systems.  In AFE's  the applied electric field causes 
$T_c$ to decrease (Fig. 1c-f, SI Fig. 2),  whereas ferroelectric materials  show the opposite tendency. 
A common feature of all studied AFE PZT   is  a negative-to-positive 
crossover that varies with temperature and composition.
Positive values of the EC $\Delta T$ are related to the reduction of isothermal entropy.
In classical FE's this is related to the drop of macroscopic polarisation  with rising temperature. 
However, in AFE's  the polarisation  may exhibit an opposite behaviour, 
i.e. increasing  with rising temperature under applied field. 
This occurs simply because the applied field favours net polarisation 
and dielectric susceptibility then increases with temperature.
The latter  results in negative change of isothermal entropy, 
and reverse electrocaloric effect  (See SI).

%%%%%%%%%%%%%%%%
%%%      PZO
%%%%%%%%%%%%%%%%
PZO  does not exhibit a macroscopic polarisation at zero  field, as expected for an AFE. 
Applied electric field induces a polarisation  that  increases up to the critical temperature, 
$T_c$, and then falls with further temperature rise (see  Fig. ~1 e, f). 
However, the induced polar state of PZO  at an applied field of 
100 MV/m  is only 18  $\mu$ C/m$^2$, which is 40 \% lower  than that of  PTO.
%
%%%%%%------REC TEXT-----%%%%%%
Ferroelectrics can also show negative ECE originated from polarisation rotation, where the polarisation 
along the field  direction increases with temperature due to approaching phase transition ~\cite{WuCohen2017a, WuCohen2017b}
Calculated behaviour of $\Delta T$ for  PZO exhibits a crossover from negative 
to positive values in the vicinity of $T_c$ as shown in Fig. 1f. 
For  applied electric fields $<$50 MV/m the EC change in temperature  exhibits 
negative values below $T_c$  
(at T=250 K  the values of ECE are  -0.7 K with applied field of 25 MV/m ). 
 
%%%%%%%%%%%%%%%%%
%%%%%%   AFE  5, 10, 15, 20
%%%%%%%%%%%%%%%%%
At zero applied field  AFE PZT ($x$=0.95, 0.9, 0.85, 0.8) shows zero macroscopic polarisation, 
but local dipoles, as will be shown  later, form competing AFE and FE domains. 
Application of an electric field enhances the polarisation, 
which reaches its maximum at  temperatures of 400 K, 350 K and 300 K characteristic 
for each composition with $x$=0.85, 0.90, 0.95, respectively (Fig. 1c, d, SI Fig. 2b, c, d).

The electrocaloric response of studied AFE's  is characterised by the negative-to-positive crossover.
In PZO  the EC $\Delta T$ changes its sign once, whereas AFE PZT exhibits more complex EC behaviour. 

%\section{Discussion}
We have found that, in general,  the  EC effect in FE and AFE \PZT~ with $x>0$ is  smaller compared to the pure FE PTO 
%%%%---new line--%%%%
(22.01 K at 100 MV/m of applied field (See SI Fig. 2a)),
%%%%%---REC TEXT----%%%%%
but at lower temperatures and, thus, more  usable under ordinary conditions.
At the AFE-FE boundary  an enhanced caloric response comparable to MPB PZT. 
The smallest EC response is observed in the pure AE PZO  of about 5 K at  100 MV/m of applied field.
The AFE PZT with $x=0.8$ exhibits the EC effect of 6.1 K at a similar field (See SI, Fig. 1). 
Meanwhile, PZT with $x$=0.95, 0.9 exhibit values of EC $\Delta T$  of about 10 K (SI Fig. 2), 
which is comparable with the EC response of  MPB PZT at similar stimuli. 

To understand the origin of the giant EC effect and negative-to-positive crossover  at the AFE-FE phase boundary
we analysed the evolution of  local dipoles in response to applied fields.  
We found that an AFE system may adopt complex dipole arrangements  
with a variety of possible states, such as dipole FE order, dipole disorder, and
various AFE dipole arrangements characterised by  zero macroscopic  
polarisation.% (as shown in Fig. 3a, b, where the system maintained 2$\times$1, 1$\times$1 patterns). 

%%%%%%%%%%%%%%%%
%%%% PZT5
%%%%%%%%%%%%%%%%
In particular, at small applied fields and low temperatures the AFE \PZTfive~  exhibits a 
dynamically stable 2$\times$1 pattern (Fig. 2a). 
Here, the local dipoles are arranged as antiparallel double pairs along $X$ cartesian direction, 
and single antiparallel arrangement along $Z$ axis (See directing arrows in the  inset of Fig. 2a).
At higher temperatures the order of local dipoles  changes to a  1$\times$1  pattern, 
where single antiparallel dipoles are alternating with the sites of dipole disorder (Fig. 2b).
Increasing the applied field to the critical value of 25 MV/m leads to the rotation of 
local dipoles, so the system turns into an induced polar FE state.  

%%%%%%%%%%%%%%%%
%%%% PZT10
%%%%%%%%%%%%%%%%
Increasing the Ti content leads to stabilisation 
of  a zig-zag pattern of AFE local dipoles  (Fig. 2c). Here, the local dipoles are arranged into  antiparallel pairs. 
As the field increases the system develops competing AFE and FE domains, with widths which   
 correlate with the strength of applied field. The critical field of  50 MV/m  switches the system 
to an induced polar FE monodomain state. 
 
%%%%%%%%%%%%%%%%
%%%% PZT15
%%%%%%%%%%%%%%%%
Higher Ti content in PbZr$_{0.85}$Ti$_{0.15}$O$_3$ increases the correlation 
length of the material,\cite{Rica2016} which leads to the formation  of stripe  ordering, with 
AFE dipole arrangement alternating with FE stripes (Fig. 2d). 

%%%%%%%% DIPOLE PATTERN %%%%%%%%%
  \begin{figure}[h]
  \centering
  \includegraphics[clip=true,width=0.48 \textwidth]{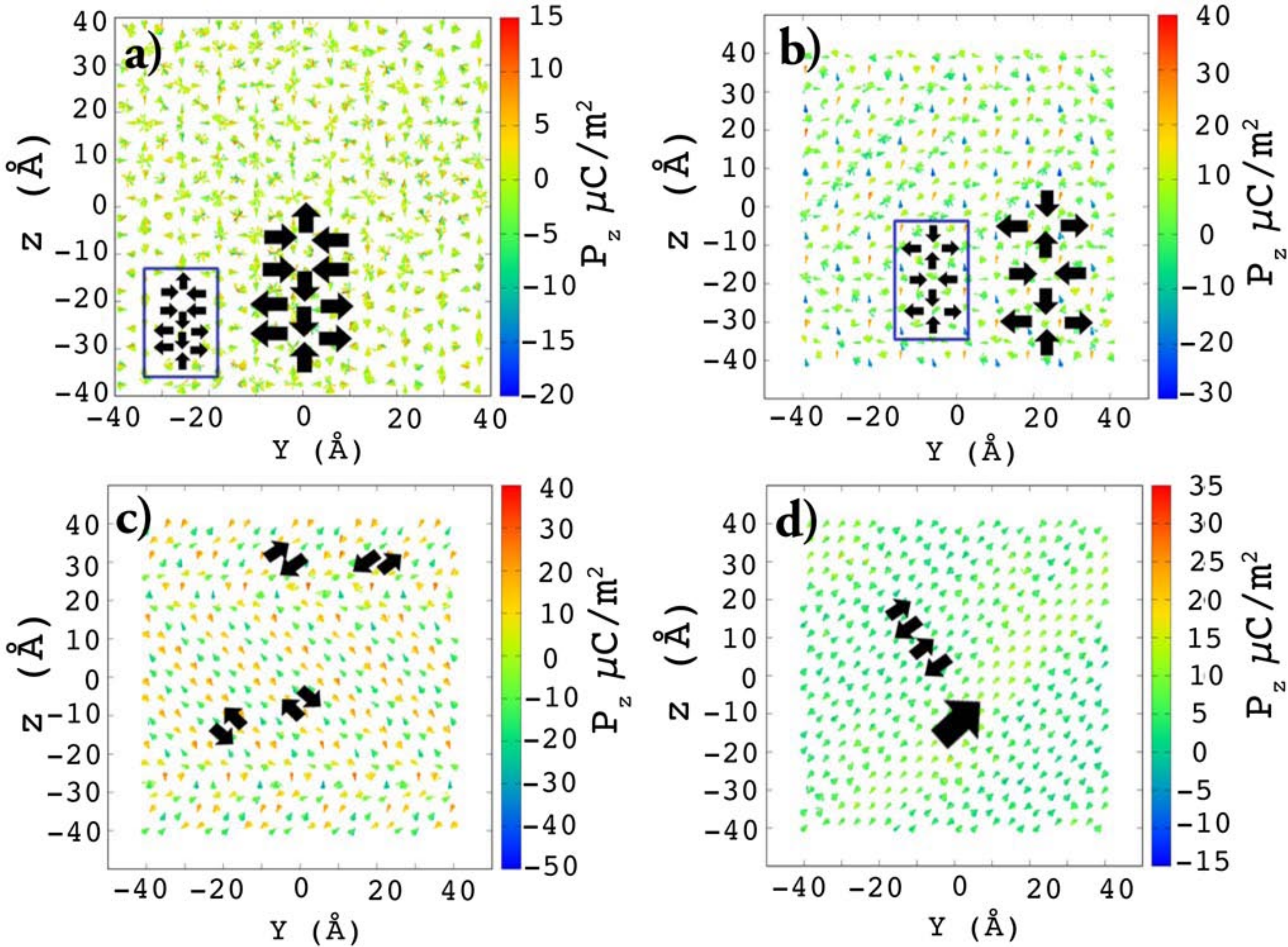}
    \caption{Gradient colour plot of projected local dipoles on XZ plane:
   a) AFE pattern  in \PZTfive~ at T=100 K and zero field. Selected area exhibits $2\times1$ pattern of local dipoles arrangement as shown in enlarged scheme with black arrows; 
   b)  AFE pattern in \PZTfive~  at T=300 K and applied E=5 MV/m. Selected  area exhibits  $1\times1$ arrangement of local dipoles; 
   c) stripe AFE-FE local dipoles arrangement  in \PZTfifteen~ at 100 K and E=5 MV/m;
   d) zig-zag-like AFE local dipoles arrangement in \PZTten~ at T=100 K at zero field.
   }
    \label{fig:PZT_5-10-15-DW}
  \end{figure}  

We suggest that the nature of giant EC temperature change, $\Delta T$, in AFE PZT 
 is related to the formation of competing AFE and induced FE orders that 
respond easily to  applied fields and temperature.  
In ferroelectrics the 
configurational entropy is related to the order maintained by a dipolar system.
In the absence of the applied field the change of polarisation with temperature, $\frac{\delta P}{\delta T}$, is relatively small, 
except the vicinity of the critical temperature, where this value is large. 
Thus, the maximum of EC $\Delta T$ in ferroelectrics  occurs when  the system switches from FE to PE.  

An AFE system may adopt  complex  configurations with a variety of 
possible dipole states - as dipole FE order, dipole disorder, and various AFE dipole arrangements 
with  zero macroscopic  polarisation (as shown in Fig. 3, 
where the system maintained 2$\times$1, 1$\times$1 patterns). We suggest that in AFE 
systems the change from AFE to FE order happen via 
a sequence of local minima with a partially preserved AFE order and the formation of competing AFE and FE domains.

%%%%%%%%%%%%%%%%%%%
%%%---new bit-mechanism %%%%%
We assume that at applied field the aligned dipole configurations may became more advantageous to anti-polar configurations. This leads to destabilisation of anti-aligned arrangements of local dipoles in an AFE material leading to their partial, or complete alignment, and, consequently formation of competing antiferroelectric and induced ferroelectric domains. The transition process may happen through initial canting as proposed in ref.~\onlinecite{Geng2015}, and follows complete rotation similar to the mechanism proposed for FE's  in ref.~\onlinecite{WuCohen2017b}.
%%%%%%%%%%%%%%%%%%%
%%%%%%%%%%%%%%%%%%%

In  bulk PZO  the change of polarisation $\frac{\delta P}{\delta T}$, is relatively small (SI Fig. 2), 
because our  system is free of defects, and grain boundaries and electrode contacts. 
Thus, each local dipole has to overcome a barrier for rotation.
However, in AFE \PZT~ the presence of different types of B-site cations
increases the configurational entropy of the  system, and supports multiple domain configurations. 
The Ti sites act as nucleation centres for the FE phase, facilitating fast response of local dipoles to applied electric fields.
At the AFE-FE phase boundary the concentration of Ti centres is such that there are no FE ordered regions in the absence of an applied electric field, however, the application of an external electric field gives rise to 
FE ordering, which competes with the AFE order (FIG. 3d).  
The maximum of EC $\Delta T$ in AFE composites occurs when  the system switches to an induced 
polar FE monodomain. 

%%%%%%%%%%%%%%%%%%%%%%%%%%%%%%%%%%%%%%%%%%%%%
%  conclusions
%%%%%%%%%%%%%%%%%%%%%%%%%%%%%%%%%%%%%%%%%%%%%
%\section{Conclusions}
We have studied the electrocaloric effect in PZT using molecular dynamics simulations with shell model forces fields. Our results show giant electrocaloric effects for FE PTO, in good agreement with similar calculations performed for FE LiNbO$_3$\cite{Rose2012}.
We found a crossover from negative to positive EC temperature change for all studied AFE PZT.
The crossover temperatures correlate with  composition, which we believe to be related to the correlation length increase in the material\cite{Rica2016}.
We have found that compositions close to the AFE-FE boundary of PZT exhibits an enhanced caloric response, comparable to that of MPB PZT but with the maximum EC temperature change occurring at temperatures closer to ambient temperatures. 
Our methodology allows to to investigate the details of the polarisation response at an atomistic level. Close to the AFE-FE boundary we identified complex dipole configurations, with competing FE and AFE domain patterns. We postulate that the small energy barriers associated with growing/ reducing these domains are responsible for the easy response of the polarisation to the applied field and temperature and, hence, for the enhanced calorific response.
Despite the high EC response, the critical temperature in many ferroelectric materials is considerably higher than room temperature, which substantially limits the potential for the application in solid-state devices. We have found that AFE PZT exhibits extrema of EC $\Delta T$ close to room temperature, in the range 300-400 K. In addition, solid solution \PZT~  offers great variability in critical temperatures and in ECE magnitude, which allows for compositional engineering of materials for electrocaloric applications. 
In summary, our findings suggest pathways for tuning the operating temperatures of ECE devices and find solutions for a broad range of operating conditions. 

\section*{Supplementary Information}
We show calculated heat capacity, and isothermal change of entropy in AFE and FE \PZT. We also calculated electrocaloric effect in  FE PbTiO$_3$, FE \PZTth, together with polarisation and electrocaloric temperature change in AFE \PZTfive, and AFE \PZTfifteen.

%\section{Acknowledgements}
Authors acknowledge UCL computational facilities LEGION and MYRIAD. 
AK is supported by the European Union's Horizon 2020 research and innovation programme under the Marie Sk$\l$odowska-Curie 
grant agreement No 796781.
REC was supported by the U. S. Office of Naval Research Grants 
No. N00014-12-1-1038, N00014-14-1-0516, and N00014-17-1-2768, 
the Carnegie Institution for Science, and the European
Research Council Advanced Grant ToMCaT.
\bibliography{ECE-revision.bib}

%merlin.mbs aipnum4-1.bst 2010-07-25 4.21a (PWD, AO, DPC) hacked
%Control: key (0)
%Control: author (8) initials jnrlst
%Control: editor formatted (1) identically to author
%Control: production of article title (0) allowed
%Control: page (1) range
%Control: year (1) truncated
%Control: production of eprint (0) enabled
\begin{thebibliography}{24}%
\makeatletter
\providecommand \@ifxundefined [1]{%
 \@ifx{#1\undefined}
}%
\providecommand \@ifnum [1]{%
 \ifnum #1\expandafter \@firstoftwo
 \else \expandafter \@secondoftwo
 \fi
}%
\providecommand \@ifx [1]{%
 \ifx #1\expandafter \@firstoftwo
 \else \expandafter \@secondoftwo
 \fi
}%
\providecommand \natexlab [1]{#1}%
\providecommand \enquote  [1]{``#1''}%
\providecommand \bibnamefont  [1]{#1}%
\providecommand \bibfnamefont [1]{#1}%
\providecommand \citenamefont [1]{#1}%
\providecommand \href@noop [0]{\@secondoftwo}%
\providecommand \href [0]{\begingroup \@sanitize@url \@href}%
\providecommand \@href[1]{\@@startlink{#1}\@@href}%
\providecommand \@@href[1]{\endgroup#1\@@endlink}%
\providecommand \@sanitize@url [0]{\catcode `\\12\catcode `\$12\catcode
  `\&12\catcode `\#12\catcode `\^12\catcode `\_12\catcode `\%12\relax}%
\providecommand \@@startlink[1]{}%
\providecommand \@@endlink[0]{}%
\providecommand \url  [0]{\begingroup\@sanitize@url \@url }%
\providecommand \@url [1]{\endgroup\@href {#1}{\urlprefix }}%
\providecommand \urlprefix  [0]{URL }%
\providecommand \Eprint [0]{\href }%
\providecommand \doibase [0]{http://dx.doi.org/}%
\providecommand \selectlanguage [0]{\@gobble}%
\providecommand \bibinfo  [0]{\@secondoftwo}%
\providecommand \bibfield  [0]{\@secondoftwo}%
\providecommand \translation [1]{[#1]}%
\providecommand \BibitemOpen [0]{}%
\providecommand \bibitemStop [0]{}%
\providecommand \bibitemNoStop [0]{.\EOS\space}%
\providecommand \EOS [0]{\spacefactor3000\relax}%
\providecommand \BibitemShut  [1]{\csname bibitem#1\endcsname}%
\let\auto@bib@innerbib\@empty
%</preamble>
\bibitem [{\citenamefont {Mischenko}\ \emph {et~al.}(2006)\citenamefont
  {Mischenko}, \citenamefont {Zhang}, \citenamefont {Scott}, \citenamefont
  {Whatmore},\ and\ \citenamefont {Mathur}}]{Mischenko2006}%
  \BibitemOpen
  \bibfield  {author} {\bibinfo {author} {\bibfnamefont {A.~S.}\ \bibnamefont
  {Mischenko}}, \bibinfo {author} {\bibfnamefont {Q.}~\bibnamefont {Zhang}},
  \bibinfo {author} {\bibfnamefont {J.~F.}\ \bibnamefont {Scott}}, \bibinfo
  {author} {\bibfnamefont {R.~W.}\ \bibnamefont {Whatmore}}, \ and\ \bibinfo
  {author} {\bibfnamefont {N.~D.}\ \bibnamefont {Mathur}},\ }\bibfield  {title}
  {\enquote {\bibinfo {title} {{Giant electrocaloric effect in thin-film
  PZT}},}\ }\href@noop {} {\bibfield  {journal} {\bibinfo  {journal} {Science}\
  }\textbf {\bibinfo {volume} {311}},\ \bibinfo {pages} {1270--1271} (\bibinfo
  {year} {2006})},\ \Eprint {http://arxiv.org/abs/0511487} {0511487}
  \BibitemShut {NoStop}%
\bibitem [{\citenamefont {Kar-Narayan}\ and\ \citenamefont
  {Mathur}(2010)}]{Kar-Narayan2010}%
  \BibitemOpen
  \bibfield  {author} {\bibinfo {author} {\bibfnamefont {S.}~\bibnamefont
  {Kar-Narayan}}\ and\ \bibinfo {author} {\bibfnamefont {N.~D.}\ \bibnamefont
  {Mathur}},\ }\bibfield  {title} {\enquote {\bibinfo {title} {{Direct and
  indirect electrocaloric measurements using multilayer capacitors}},}\ }\href
  {http://arxiv.org/abs/0912.1978} {\bibfield  {journal} {\bibinfo  {journal}
  {Journal of Physics D: Applied Physics}\ }\textbf {\bibinfo {volume} {43}},\
  \bibinfo {pages} {032002} (\bibinfo {year} {2010})},\ \Eprint
  {http://arxiv.org/abs/0912.1978} {0912.1978} \BibitemShut {NoStop}%
\bibitem [{\citenamefont {Lu}\ \emph {et~al.}(2010)\citenamefont {Lu},
  \citenamefont {Ro{\v{z}}i{\v{c}}}, \citenamefont {Zhang}, \citenamefont
  {Kutnjak}, \citenamefont {Li}, \citenamefont {Furman}, \citenamefont {Gorny},
  \citenamefont {Lin}, \citenamefont {Mali{\v{c}}}, \citenamefont {Kosec},
  \citenamefont {Blinc},\ and\ \citenamefont {Pirc}}]{Lu2010}%
  \BibitemOpen
  \bibfield  {author} {\bibinfo {author} {\bibfnamefont {S.~G.}\ \bibnamefont
  {Lu}}, \bibinfo {author} {\bibfnamefont {B.}~\bibnamefont
  {Ro{\v{z}}i{\v{c}}}}, \bibinfo {author} {\bibfnamefont {Q.~M.}\ \bibnamefont
  {Zhang}}, \bibinfo {author} {\bibfnamefont {Z.}~\bibnamefont {Kutnjak}},
  \bibinfo {author} {\bibfnamefont {X.}~\bibnamefont {Li}}, \bibinfo {author}
  {\bibfnamefont {E.}~\bibnamefont {Furman}}, \bibinfo {author} {\bibfnamefont
  {L.~J.}\ \bibnamefont {Gorny}}, \bibinfo {author} {\bibfnamefont
  {M.}~\bibnamefont {Lin}}, \bibinfo {author} {\bibfnamefont {B.}~\bibnamefont
  {Mali{\v{c}}}}, \bibinfo {author} {\bibfnamefont {M.}~\bibnamefont {Kosec}},
  \bibinfo {author} {\bibfnamefont {R.}~\bibnamefont {Blinc}}, \ and\ \bibinfo
  {author} {\bibfnamefont {R.}~\bibnamefont {Pirc}},\ }\bibfield  {title}
  {\enquote {\bibinfo {title} {{Organic and inorganic relaxor ferroelectrics
  with giant electrocaloric effect}},}\ }\href {\doibase 10.1063/1.3501975}
  {\bibfield  {journal} {\bibinfo  {journal} {Applied Physics Letters}\
  }\textbf {\bibinfo {volume} {97}},\ \bibinfo {pages} {16} (\bibinfo {year}
  {2010})}\BibitemShut {NoStop}%
\bibitem [{\citenamefont {Woodward}, \citenamefont {Knudsen},\ and\
  \citenamefont {Reaney}(2005)}]{Woodward2005}%
  \BibitemOpen
  \bibfield  {author} {\bibinfo {author} {\bibfnamefont {D.~I.}\ \bibnamefont
  {Woodward}}, \bibinfo {author} {\bibfnamefont {J.}~\bibnamefont {Knudsen}}, \
  and\ \bibinfo {author} {\bibfnamefont {I.~M.}\ \bibnamefont {Reaney}},\
  }\bibfield  {title} {\enquote {\bibinfo {title} {{Review of crystal and
  domain structures in the PZT solid solution}},}\ }\href {\doibase
  10.1103/PhysRevB.72.104110} {\bibfield  {journal} {\bibinfo  {journal}
  {Physical Review B}\ }\textbf {\bibinfo {volume} {72}},\ \bibinfo {pages}
  {104110} (\bibinfo {year} {2005})}\BibitemShut {NoStop}%
\bibitem [{\citenamefont {Zhang}\ \emph {et~al.}(2016)\citenamefont {Zhang},
  \citenamefont {Li}, \citenamefont {Hou}, \citenamefont {Yu}, \citenamefont
  {Cao}, \citenamefont {Feng},\ and\ \citenamefont {Fei}}]{Zhang2016}%
  \BibitemOpen
  \bibfield  {author} {\bibinfo {author} {\bibfnamefont {T.}~\bibnamefont
  {Zhang}}, \bibinfo {author} {\bibfnamefont {W.}~\bibnamefont {Li}}, \bibinfo
  {author} {\bibfnamefont {Y.-F.}\ \bibnamefont {Hou}}, \bibinfo {author}
  {\bibfnamefont {Y.}~\bibnamefont {Yu}}, \bibinfo {author} {\bibfnamefont
  {W.~P.}\ \bibnamefont {Cao}}, \bibinfo {author} {\bibfnamefont
  {Y.}~\bibnamefont {Feng}}, \ and\ \bibinfo {author} {\bibfnamefont
  {W.}~\bibnamefont {Fei}},\ }\bibfield  {title} {\enquote {\bibinfo {title}
  {{Positive/Negative Electrocaloric Effect Induced by Defect Dipoles in PZT
  Ferroelectric Bilayer Thin Films}},}\ }\href {\doibase 10.1039/C6RA14776C}
  {\bibfield  {journal} {\bibinfo  {journal} {Royal Society of Chemistry
  Advances}\ }\textbf {\bibinfo {volume} {6}},\ \bibinfo {pages} {71934--71939}
  (\bibinfo {year} {2016})}\BibitemShut {NoStop}%
\bibitem [{\citenamefont {Zuo}\ \emph {et~al.}(2015)\citenamefont {Zuo},
  \citenamefont {Chen}, \citenamefont {Wang}, \citenamefont {Yang},
  \citenamefont {Zhan}, \citenamefont {Liu}, \citenamefont {Wang},\ and\
  \citenamefont {Li}}]{Zuo2015}%
  \BibitemOpen
  \bibfield  {author} {\bibinfo {author} {\bibfnamefont {Z.}~\bibnamefont
  {Zuo}}, \bibinfo {author} {\bibfnamefont {B.}~\bibnamefont {Chen}}, \bibinfo
  {author} {\bibfnamefont {B.}~\bibnamefont {Wang}}, \bibinfo {author}
  {\bibfnamefont {H.}~\bibnamefont {Yang}}, \bibinfo {author} {\bibfnamefont
  {Q.}~\bibnamefont {Zhan}}, \bibinfo {author} {\bibfnamefont {Y.}~\bibnamefont
  {Liu}}, \bibinfo {author} {\bibfnamefont {J.}~\bibnamefont {Wang}}, \ and\
  \bibinfo {author} {\bibfnamefont {R.-W.}\ \bibnamefont {Li}},\ }\bibfield
  {title} {\enquote {\bibinfo {title} {{Strain assisted electrocaloric effect
  in Pb(Zr$_{0.95}$Ti$_{0.05}$)O$_3$ films on
  0.7Pb(Mg$_{1/3}$Nb$_{2/3}$)O$_3$-0.3PbTiO$_3$ substrate}},}\ }\href
  {http://www.nature.com/articles/srep16164} {\bibfield  {journal} {\bibinfo
  {journal} {Scientific Reports}\ }\textbf {\bibinfo {volume} {5}},\ \bibinfo
  {pages} {16164} (\bibinfo {year} {2015})}\BibitemShut {NoStop}%
\bibitem [{\citenamefont {Geng}\ \emph {et~al.}(2015)\citenamefont {Geng},
  \citenamefont {Y.~Liu}, \citenamefont {Bellaiche}, \citenamefont {Scott},
  \citenamefont {Dkhil},\ and\ \citenamefont {Jiang}}]{Geng2015}%
  \BibitemOpen
  \bibfield  {author} {\bibinfo {author} {\bibfnamefont {W.}~\bibnamefont
  {Geng}}, \bibinfo {author} {\bibfnamefont {X.~M.}\ \bibnamefont {Y.~Liu}},
  \bibinfo {author} {\bibfnamefont {L.}~\bibnamefont {Bellaiche}}, \bibinfo
  {author} {\bibfnamefont {J.~F.}\ \bibnamefont {Scott}}, \bibinfo {author}
  {\bibfnamefont {B.}~\bibnamefont {Dkhil}}, \ and\ \bibinfo {author}
  {\bibfnamefont {A.}~\bibnamefont {Jiang}},\ }\bibfield  {title} {\enquote
  {\bibinfo {title} {{ Giant Negative Electrocaloric Effect in
  Antiferroelectric La-Doped Pb(ZrTi)O$_3$ Thin Films Near Room
  Temperature}},}\ }\href@noop {} {\bibfield  {journal} {\bibinfo  {journal}
  {Advanced Materials}\ }\textbf {\bibinfo {volume} {27(20)}},\ \bibinfo
  {pages} {3165} (\bibinfo {year} {2015})}\BibitemShut {NoStop}%
\bibitem [{\citenamefont {Kingsland}, \citenamefont {Lisenkov},\ and\
  \citenamefont {Ponomareva}(2018)}]{Ponomareva2018}%
  \BibitemOpen
  \bibfield  {author} {\bibinfo {author} {\bibfnamefont {M.}~\bibnamefont
  {Kingsland}}, \bibinfo {author} {\bibfnamefont {S.}~\bibnamefont {Lisenkov}},
  \ and\ \bibinfo {author} {\bibfnamefont {I.}~\bibnamefont {Ponomareva}},\
  }\bibfield  {title} {\enquote {\bibinfo {title} {Unveiling electrocaloric
  potential of antiferroelectrics with phase competition},}\ }\href@noop {}
  {\bibfield  {journal} {\bibinfo  {journal} {Advanced Theory Simulations}\
  }\textbf {\bibinfo {volume} {1}},\ \bibinfo {pages} {1800096} (\bibinfo
  {year} {2018})}\BibitemShut {NoStop}%
\bibitem [{\citenamefont {Pirc}\ \emph {et~al.}(2014)\citenamefont {Pirc},
  \citenamefont {Ro{\v{z}}i{\v{c}}}, \citenamefont {Koruza}, \citenamefont
  {Mali{\v{c}}},\ and\ \citenamefont {Kutnjak}}]{Pirc2014}%
  \BibitemOpen
  \bibfield  {author} {\bibinfo {author} {\bibfnamefont {R.}~\bibnamefont
  {Pirc}}, \bibinfo {author} {\bibfnamefont {B.}~\bibnamefont
  {Ro{\v{z}}i{\v{c}}}}, \bibinfo {author} {\bibfnamefont {J.}~\bibnamefont
  {Koruza}}, \bibinfo {author} {\bibfnamefont {B.}~\bibnamefont {Mali{\v{c}}}},
  \ and\ \bibinfo {author} {\bibfnamefont {Z.}~\bibnamefont {Kutnjak}},\
  }\bibfield  {title} {\enquote {\bibinfo {title} {{Negative electrocaloric
  effect in antiferroelectric PbZrO$_3$}},}\ }\href
  {http://stacks.iop.org/0295-5075/107/i=1/a=17002?key=crossref.538a71d91bea5e0482d3d27fb7e2038e}
  {\bibfield  {journal} {\bibinfo  {journal} {Europhysics Letters}\ }\textbf
  {\bibinfo {volume} {107}},\ \bibinfo {pages} {17002} (\bibinfo {year}
  {2014})}\BibitemShut {NoStop}%
\bibitem [{\citenamefont {Rose}\ and\ \citenamefont {Cohen}(2012)}]{Rose2012}%
  \BibitemOpen
  \bibfield  {author} {\bibinfo {author} {\bibfnamefont {M.~C.}\ \bibnamefont
  {Rose}}\ and\ \bibinfo {author} {\bibfnamefont {R.~E.}\ \bibnamefont
  {Cohen}},\ }\bibfield  {title} {\enquote {\bibinfo {title} {{Giant
  electrocaloric effect around $T_c$}},}\ }\href {\doibase
  10.1103/PhysRevLett.109.187604} {\bibfield  {journal} {\bibinfo  {journal}
  {Physical Review Letters}\ }\textbf {\bibinfo {volume} {109}},\ \bibinfo
  {pages} {1} (\bibinfo {year} {2012})}\BibitemShut {NoStop}%
\bibitem [{\citenamefont {Rose}\ and\ \citenamefont {Cohen}(2014)}]{Erratum14}%
  \BibitemOpen
  \bibfield  {author} {\bibinfo {author} {\bibfnamefont {M.}~\bibnamefont
  {Rose}}\ and\ \bibinfo {author} {\bibfnamefont {R.~E.}\ \bibnamefont
  {Cohen}},\ }\bibfield  {title} {\enquote {\bibinfo {title} {{Erratum: Giant
  Electrocaloric Effect Around Tc}},}\ }\href@noop {} {\bibfield  {journal}
  {\bibinfo  {journal} {Phys. Rev. Lett.}\ }\textbf {\bibinfo {volume} {112}},\
  \bibinfo {pages} {249901} (\bibinfo {year} {2014})}\BibitemShut {NoStop}%
\bibitem [{\citenamefont {Wu}\ and\ \citenamefont
  {Cohen}(2017{\natexlab{a}})}]{WuCohen2017a}%
  \BibitemOpen
  \bibfield  {author} {\bibinfo {author} {\bibfnamefont {H.~H.}\ \bibnamefont
  {Wu}}\ and\ \bibinfo {author} {\bibfnamefont {R.~E.}\ \bibnamefont {Cohen}},\
  }\bibfield  {title} {\enquote {\bibinfo {title} {Electric-field-induced phase
  transition and electrocaloric effect in \ce{PMN-PT}},}\ }\href {\doibase ARTN
  054116 10.1103/PhysRevB.96.054116} {\bibfield  {journal} {\bibinfo  {journal}
  {Physical Review B}\ }\textbf {\bibinfo {volume} {96}},\ \bibinfo {pages}
  {054116} (\bibinfo {year} {2017}{\natexlab{a}})}\BibitemShut {NoStop}%
\bibitem [{\citenamefont {Wu}\ and\ \citenamefont
  {Cohen}(2017{\natexlab{b}})}]{WuCohen2017b}%
  \BibitemOpen
  \bibfield  {author} {\bibinfo {author} {\bibfnamefont {H.~H.}\ \bibnamefont
  {Wu}}\ and\ \bibinfo {author} {\bibfnamefont {R.~E.}\ \bibnamefont {Cohen}},\
  }\bibfield  {title} {\enquote {\bibinfo {title} {Polarization rotation and
  the electrocaloric effect in barium titanate},}\ }\href {\doibase
  10.1088/1361-648X/aa94db} {\bibfield  {journal} {\bibinfo  {journal} {Journal
  of Physics: Condensed Matter}\ }\textbf {\bibinfo {volume} {29}},\ \bibinfo
  {pages} {485704} (\bibinfo {year} {2017}{\natexlab{b}})}\BibitemShut
  {NoStop}%
\bibitem [{\citenamefont {Gindele}\ \emph {et~al.}(2015)\citenamefont
  {Gindele}, \citenamefont {Kimmel}, \citenamefont {Cain},\ and\ \citenamefont
  {Duffy}}]{Gindele2015}%
  \BibitemOpen
  \bibfield  {author} {\bibinfo {author} {\bibfnamefont {O.}~\bibnamefont
  {Gindele}}, \bibinfo {author} {\bibfnamefont {A.}~\bibnamefont {Kimmel}},
  \bibinfo {author} {\bibfnamefont {M.~G.}\ \bibnamefont {Cain}}, \ and\
  \bibinfo {author} {\bibfnamefont {D.}~\bibnamefont {Duffy}},\ }\bibfield
  {title} {\enquote {\bibinfo {title} {{Shell Model force field for Lead
  Zirconate Titanate Pb(Zr$_{1- x}$Ti$_x$)O$_3$ }},}\ }\href {\doibase
  10.1021/acs.jpcc.5b03207} {\bibfield  {journal} {\bibinfo  {journal} {Journal
  of Physical Chemistry C}\ }\textbf {\bibinfo {volume} {119}},\ \bibinfo
  {pages} {17784--17789} (\bibinfo {year} {2015})}\BibitemShut {NoStop}%
\bibitem [{\citenamefont {Todorov}\ \emph {et~al.}(2006)\citenamefont
  {Todorov}, \citenamefont {Smith}, \citenamefont {Trachenko},\ and\
  \citenamefont {Dove}}]{Todorov2006}%
  \BibitemOpen
  \bibfield  {author} {\bibinfo {author} {\bibfnamefont {I.~T.}\ \bibnamefont
  {Todorov}}, \bibinfo {author} {\bibfnamefont {W.}~\bibnamefont {Smith}},
  \bibinfo {author} {\bibfnamefont {K.}~\bibnamefont {Trachenko}}, \ and\
  \bibinfo {author} {\bibfnamefont {M.~T.}\ \bibnamefont {Dove}},\ }\bibfield
  {title} {\enquote {\bibinfo {title} {{DL{\_}POLY{\_}3: new dimensions in
  molecular dynamics simulations via massive parallelism}},}\ }\href {\doibase
  10.1039/b517931a} {\bibfield  {journal} {\bibinfo  {journal} {Journal of
  Materials Chemistry}\ }\textbf {\bibinfo {volume} {16}},\ \bibinfo {pages}
  {1911} (\bibinfo {year} {2006})}\BibitemShut {NoStop}%
\bibitem [{\citenamefont {Fincham}\ and\ \citenamefont
  {Mitchell}(1993)}]{Fincham93}%
  \BibitemOpen
  \bibfield  {author} {\bibinfo {author} {\bibfnamefont {D.}~\bibnamefont
  {Fincham}}\ and\ \bibinfo {author} {\bibfnamefont {P.~J.}\ \bibnamefont
  {Mitchell}},\ }\bibfield  {title} {\enquote {\bibinfo {title} {{Shell model
  simulations by adiabatic dynamics}},}\ }\href@noop {} {\bibfield  {journal}
  {\bibinfo  {journal} {J. Phys. Condens. Matter}\ }\textbf {\bibinfo {volume}
  {5}},\ \bibinfo {pages} {1031} (\bibinfo {year} {1993})}\BibitemShut
  {NoStop}%
\bibitem [{\citenamefont {Morimoto}\ \emph {et~al.}(2003)\citenamefont
  {Morimoto}, \citenamefont {Uematsu}, \citenamefont {Sawai}, \citenamefont
  {Hisano},\ and\ \citenamefont {Yamamoto}}]{Morimoto2003}%
  \BibitemOpen
  \bibfield  {author} {\bibinfo {author} {\bibfnamefont {K.}~\bibnamefont
  {Morimoto}}, \bibinfo {author} {\bibfnamefont {A.}~\bibnamefont {Uematsu}},
  \bibinfo {author} {\bibfnamefont {S.}~\bibnamefont {Sawai}}, \bibinfo
  {author} {\bibfnamefont {K.}~\bibnamefont {Hisano}}, \ and\ \bibinfo {author}
  {\bibfnamefont {T.}~\bibnamefont {Yamamoto}},\ }\bibfield  {title} {\enquote
  {\bibinfo {title} {{Simultaneous Measurement of Thermophysical Properties and
  Dielectric Properties of PZT-Based Ferroelectric Ceramics by Thermal
  Radiation Calorimetry}},}\ }\href@noop {} {\bibfield  {journal} {\bibinfo
  {journal} {International Journal of Thermophysics}\ }\textbf {\bibinfo
  {volume} {24}},\ \bibinfo {pages} {3} (\bibinfo {year} {2003})}\BibitemShut
  {NoStop}%
\bibitem [{\citenamefont {Noheda}\ \emph {et~al.}(1999)\citenamefont {Noheda},
  \citenamefont {Cox}, \citenamefont {Shirane}, \citenamefont {Gonzalo},
  \citenamefont {Cross},\ and\ \citenamefont {Park}}]{Noheda1999}%
  \BibitemOpen
  \bibfield  {author} {\bibinfo {author} {\bibfnamefont {B.}~\bibnamefont
  {Noheda}}, \bibinfo {author} {\bibfnamefont {D.~E.}\ \bibnamefont {Cox}},
  \bibinfo {author} {\bibfnamefont {G.}~\bibnamefont {Shirane}}, \bibinfo
  {author} {\bibfnamefont {J.~A.}\ \bibnamefont {Gonzalo}}, \bibinfo {author}
  {\bibfnamefont {L.~E.}\ \bibnamefont {Cross}}, \ and\ \bibinfo {author}
  {\bibfnamefont {S.-E.}\ \bibnamefont {Park}},\ }\bibfield  {title} {\enquote
  {\bibinfo {title} {{A monoclinic ferroelectric phase in the
  Pb(Zr$_{(1-x)}$Ti$_x$)O$_3$ solid solution}},}\ }\href {\doibase
  10.1063/1.123756} {\bibfield  {journal} {\bibinfo  {journal} {Applied Physics
  Letters}\ }\textbf {\bibinfo {volume} {2059}},\ \bibinfo {pages} {6}
  (\bibinfo {year} {1999})},\ \Eprint {http://arxiv.org/abs/9903007} {9903007}
  \BibitemShut {NoStop}%
\bibitem [{\citenamefont {Cohen}(2018)}]{Cohen2018}%
  \BibitemOpen
  \bibfield  {author} {\bibinfo {author} {\bibfnamefont {R.~E.}\ \bibnamefont
  {Cohen}},\ }\bibfield  {title} {\enquote {\bibinfo {title} {{Morphing into
  action}},}\ }\href@noop {} {\bibfield  {journal} {\bibinfo  {journal}
  {Nature}\ }\textbf {\bibinfo {volume} {562}},\ \bibinfo {pages} {48--49}
  (\bibinfo {year} {2018})}\BibitemShut {NoStop}%
\bibitem [{\citenamefont {Zhang}\ \emph {et~al.}(2014)\citenamefont {Zhang},
  \citenamefont {Yokota}, \citenamefont {Glazer}, \citenamefont {Ren},
  \citenamefont {Keen}, \citenamefont {A.~Keeble}, \citenamefont {Thomas},\
  and\ \citenamefont {Ye}}]{Glazer2014}%
  \BibitemOpen
  \bibfield  {author} {\bibinfo {author} {\bibfnamefont {N.}~\bibnamefont
  {Zhang}}, \bibinfo {author} {\bibfnamefont {H.}~\bibnamefont {Yokota}},
  \bibinfo {author} {\bibfnamefont {A.~M.}\ \bibnamefont {Glazer}}, \bibinfo
  {author} {\bibfnamefont {Z.}~\bibnamefont {Ren}}, \bibinfo {author}
  {\bibfnamefont {D.}~\bibnamefont {Keen}}, \bibinfo {author} {\bibfnamefont
  {D.~S.}\ \bibnamefont {A.~Keeble}}, \bibinfo {author} {\bibfnamefont {P.~A.}\
  \bibnamefont {Thomas}}, \ and\ \bibinfo {author} {\bibfnamefont {Z.-G.}\
  \bibnamefont {Ye}},\ }\bibfield  {title} {\enquote {\bibinfo {title} {{The
  missing boundary in the phase diagram of PbZrTiO$_3$}},}\ }\href {\doibase
  10.1038/ncomms6231} {\bibfield  {journal} {\bibinfo  {journal} {Nature
  Communication}\ }\textbf {\bibinfo {volume} {5}},\ \bibinfo {pages} {5231}
  (\bibinfo {year} {2014})}\BibitemShut {NoStop}%
\bibitem [{\citenamefont {Bogdanov}\ \emph {et~al.}(2016)\citenamefont
  {Bogdanov}, \citenamefont {Mysovsky}, \citenamefont {Pickard},\ and\
  \citenamefont {Kimmel}}]{Bogdanov2016}%
  \BibitemOpen
  \bibfield  {author} {\bibinfo {author} {\bibfnamefont {A.}~\bibnamefont
  {Bogdanov}}, \bibinfo {author} {\bibfnamefont {A.}~\bibnamefont {Mysovsky}},
  \bibinfo {author} {\bibfnamefont {C.~J.}\ \bibnamefont {Pickard}}, \ and\
  \bibinfo {author} {\bibfnamefont {A.~V.}\ \bibnamefont {Kimmel}},\ }\bibfield
   {title} {\enquote {\bibinfo {title} {{Modelling the structure of Zr-rich PZT
  by a multiphase approach}},}\ }\href@noop {} {\bibfield  {journal} {\bibinfo
  {journal} {Physical Chemistry Chemical Physics}\ }\textbf {\bibinfo {volume}
  {18}},\ \bibinfo {pages} {28316} (\bibinfo {year} {2016})}\BibitemShut
  {NoStop}%
\bibitem [{\citenamefont {Fu}\ and\ \citenamefont {Cohen}(2000)}]{FuCohen2000}%
  \BibitemOpen
  \bibfield  {author} {\bibinfo {author} {\bibfnamefont {H.}~\bibnamefont
  {Fu}}\ and\ \bibinfo {author} {\bibfnamefont {R.~E.}\ \bibnamefont {Cohen}},\
  }\bibfield  {title} {\enquote {\bibinfo {title} {Polarization rotation
  mechanism for ultrahigh electromechanical response in single-crystal
  piezoelectrics},}\ }\href@noop {} {\bibfield  {journal} {\bibinfo  {journal}
  {Nature}\ }\textbf {\bibinfo {volume} {403}},\ \bibinfo {pages} {281--283}
  (\bibinfo {year} {2000})}\BibitemShut {NoStop}%
\bibitem [{\citenamefont {Ahart}\ \emph {et~al.}(2008)\citenamefont {Ahart},
  \citenamefont {Somayazulu}, \citenamefont {Cohen}, \citenamefont {Ganesh},
  \citenamefont {Dera}, \citenamefont {Mao}, \citenamefont {Hemley},
  \citenamefont {Ren}, \citenamefont {Liermann},\ and\ \citenamefont
  {Wu}}]{Ahart2008}%
  \BibitemOpen
  \bibfield  {author} {\bibinfo {author} {\bibfnamefont {M.}~\bibnamefont
  {Ahart}}, \bibinfo {author} {\bibfnamefont {M.}~\bibnamefont {Somayazulu}},
  \bibinfo {author} {\bibfnamefont {R.~E.}\ \bibnamefont {Cohen}}, \bibinfo
  {author} {\bibfnamefont {P.}~\bibnamefont {Ganesh}}, \bibinfo {author}
  {\bibfnamefont {P.}~\bibnamefont {Dera}}, \bibinfo {author} {\bibfnamefont
  {H.~K.}\ \bibnamefont {Mao}}, \bibinfo {author} {\bibfnamefont {R.~J.}\
  \bibnamefont {Hemley}}, \bibinfo {author} {\bibfnamefont {Y.}~\bibnamefont
  {Ren}}, \bibinfo {author} {\bibfnamefont {P.}~\bibnamefont {Liermann}}, \
  and\ \bibinfo {author} {\bibfnamefont {Z.}~\bibnamefont {Wu}},\ }\bibfield
  {title} {\enquote {\bibinfo {title} {Origin of morphotropic phase boundaries
  in ferroelectrics},}\ }\href@noop {} {\bibfield  {journal} {\bibinfo
  {journal} {Nature}\ }\textbf {\bibinfo {volume} {451}},\ \bibinfo {pages}
  {545} (\bibinfo {year} {2008})}\BibitemShut {NoStop}%
\bibitem [{\citenamefont {Guzman-Verri}\ and\ \citenamefont
  {Littlewood}(2016)}]{Rica2016}%
  \BibitemOpen
  \bibfield  {author} {\bibinfo {author} {\bibfnamefont {G.~G.}\ \bibnamefont
  {Guzman-Verri}}\ and\ \bibinfo {author} {\bibfnamefont {P.~B.}\ \bibnamefont
  {Littlewood}},\ }\bibfield  {title} {\enquote {\bibinfo {title} {{Why is the
  electrocaloric effect so small in ferroelectrics?}}}\ }\href {\doibase
  10.1063/1.4950788} {\bibfield  {journal} {\bibinfo  {journal} {Applied
  Physics Letters Materials}\ }\textbf {\bibinfo {volume} {4}},\ \bibinfo
  {pages} {064106} (\bibinfo {year} {2016})}\BibitemShut {NoStop}%
\end{thebibliography}%


%merlin.mbs aipnum4-1.bst 2010-07-25 4.21a (PWD, AO, DPC) hacked
%Control: key (0)
%Control: author (8) initials jnrlst
%Control: editor formatted (1) identically to author
%Control: production of article title (0) allowed
%Control: page (1) range
%Control: year (1) truncated
%Control: production of eprint (0) enabled
\begin{thebibliography}{3}%
\makeatletter
\providecommand \@ifxundefined [1]{%
 \@ifx{#1\undefined}
}%
\providecommand \@ifnum [1]{%
 \ifnum #1\expandafter \@firstoftwo
 \else \expandafter \@secondoftwo
 \fi
}%
\providecommand \@ifx [1]{%
 \ifx #1\expandafter \@firstoftwo
 \else \expandafter \@secondoftwo
 \fi
}%
\providecommand \natexlab [1]{#1}%
\providecommand \enquote  [1]{``#1''}%
\providecommand \bibnamefont  [1]{#1}%
\providecommand \bibfnamefont [1]{#1}%
\providecommand \citenamefont [1]{#1}%
\providecommand \href@noop [0]{\@secondoftwo}%
\providecommand \href [0]{\begingroup \@sanitize@url \@href}%
\providecommand \@href[1]{\@@startlink{#1}\@@href}%
\providecommand \@@href[1]{\endgroup#1\@@endlink}%
\providecommand \@sanitize@url [0]{\catcode `\\12\catcode `\$12\catcode
  `\&12\catcode `\#12\catcode `\^12\catcode `\_12\catcode `\%12\relax}%
\providecommand \@@startlink[1]{}%
\providecommand \@@endlink[0]{}%
\providecommand \url  [0]{\begingroup\@sanitize@url \@url }%
\providecommand \@url [1]{\endgroup\@href {#1}{\urlprefix }}%
\providecommand \urlprefix  [0]{URL }%
\providecommand \Eprint [0]{\href }%
\providecommand \doibase [0]{http://dx.doi.org/}%
\providecommand \selectlanguage [0]{\@gobble}%
\providecommand \bibinfo  [0]{\@secondoftwo}%
\providecommand \bibfield  [0]{\@secondoftwo}%
\providecommand \translation [1]{[#1]}%
\providecommand \BibitemOpen [0]{}%
\providecommand \bibitemStop [0]{}%
\providecommand \bibitemNoStop [0]{.\EOS\space}%
\providecommand \EOS [0]{\spacefactor3000\relax}%
\providecommand \BibitemShut  [1]{\csname bibitem#1\endcsname}%
\let\auto@bib@innerbib\@empty
%</preamble>
\bibitem [{\citenamefont {Morimoto}\ \emph {et~al.}(2003)\citenamefont
  {Morimoto}, \citenamefont {Uematsu}, \citenamefont {Sawai}, \citenamefont
  {Hisano},\ and\ \citenamefont {Yamamoto}}]{Morimoto2003}%
  \BibitemOpen
  \bibfield  {author} {\bibinfo {author} {\bibfnamefont {K.}~\bibnamefont
  {Morimoto}}, \bibinfo {author} {\bibfnamefont {A.}~\bibnamefont {Uematsu}},
  \bibinfo {author} {\bibfnamefont {S.}~\bibnamefont {Sawai}}, \bibinfo
  {author} {\bibfnamefont {K.}~\bibnamefont {Hisano}}, \ and\ \bibinfo {author}
  {\bibfnamefont {T.}~\bibnamefont {Yamamoto}},\ }\bibfield  {title} {\enquote
  {\bibinfo {title} {{Simultaneous Measurement of Thermophysical Properties and
  Dielectric Properties of PZT-Based Ferroelectric Ceramics by Thermal
  Radiation Calorimetry}},}\ }\href@noop {} {\bibfield  {journal} {\bibinfo
  {journal} {International Journal of Thermophysics}\ }\textbf {\bibinfo
  {volume} {24}},\ \bibinfo {pages} {3} (\bibinfo {year} {2003})}\BibitemShut
  {NoStop}%
\bibitem [{\citenamefont {Yoshida}\ \emph {et~al.}(2009)\citenamefont
  {Yoshida}, \citenamefont {Moriya}, \citenamefont {Tojo}, \citenamefont
  {Kawaji}, \citenamefont {Atake},\ and\ \citenamefont
  {Kuroiwa}}]{Yoshida2009}%
  \BibitemOpen
  \bibfield  {author} {\bibinfo {author} {\bibfnamefont {T.}~\bibnamefont
  {Yoshida}}, \bibinfo {author} {\bibfnamefont {Y.}~\bibnamefont {Moriya}},
  \bibinfo {author} {\bibfnamefont {T.}~\bibnamefont {Tojo}}, \bibinfo {author}
  {\bibfnamefont {H.}~\bibnamefont {Kawaji}}, \bibinfo {author} {\bibfnamefont
  {T.}~\bibnamefont {Atake}}, \ and\ \bibinfo {author} {\bibfnamefont
  {Y.}~\bibnamefont {Kuroiwa}},\ }\bibfield  {title} {\enquote {\bibinfo
  {title} {{Heat Capacity at Constant Pressure and Thermodynamic properties of
  phase transitions in PbMO$_3$ (M=Ti, Zr AND Hf)}},}\ }\href@noop {}
  {\bibfield  {journal} {\bibinfo  {journal} {Journal of Thermal Analysis and
  Calorimetry}\ }\textbf {\bibinfo {volume} {95}},\ \bibinfo {pages} {675}
  (\bibinfo {year} {2009})}\BibitemShut {NoStop}%
\bibitem [{\citenamefont {Rossetti}\ and\ \citenamefont
  {Navrotsky}(1999)}]{Rosetti1999}%
  \BibitemOpen
  \bibfield  {author} {\bibinfo {author} {\bibfnamefont {G.}~\bibnamefont
  {Rossetti}}\ and\ \bibinfo {author} {\bibfnamefont {A.}~\bibnamefont
  {Navrotsky}},\ }\bibfield  {title} {\enquote {\bibinfo {title} {{Calorimetric
  Investigation of Tricritical Behavior in Tetragonal PbZrTiO$_3$}},}\
  }\href@noop {} {\bibfield  {journal} {\bibinfo  {journal} {Journal of Solid
  State Chemistry}\ }\textbf {\bibinfo {volume} {144}},\ \bibinfo {pages} {188}
  (\bibinfo {year} {1999})}\BibitemShut {NoStop}%
\end{thebibliography}%
\end{document}

\preprint{AIP/123-QED}

\title{Supplementary Information\\ Giant electrocaloric effect at the antiferroelecrtric-to-ferroelectric  phase boundary in \PZT  }

\maketitle

 %%%%%%%%%%%%%%%%%%%%%%%%%%%%%%%%%%% 
%   Cp calculated
%%%%%%%%%%%%%%%%%%%%%%%%%%%%%%%%%%%  
\subsection{Calculated specific heat capacity} 

The values of specific heat capacity, $C_{p,E}$, for studied PZT systems were calculated as the derivative of the 
total energy with temperature ($\frac{\partial E_{tot}}{\partial T}$) 
at a given value of electric field, $E$,  estimated as the slope of the linear fit to the total energy (E$_{tot}$) versus $T$. 
The calculated $C_{p,E}$ values show good agreement with the Dulong-Petit limit of 3$k_B$/atom for crystals. Our core-shell force field systematically underestimates the Curie temperature, thus,  the peaks of heat capacity curves are shifted towards lower temperatures with respect to the experimental values ~\cite{Morimoto2003, Yoshida2009, Rosetti1999}.

%%%%%%  FIG1 %%%%%%
 \begin{figure}[h]
  \centering
  \includegraphics[clip=true,width=0.45\textwidth]{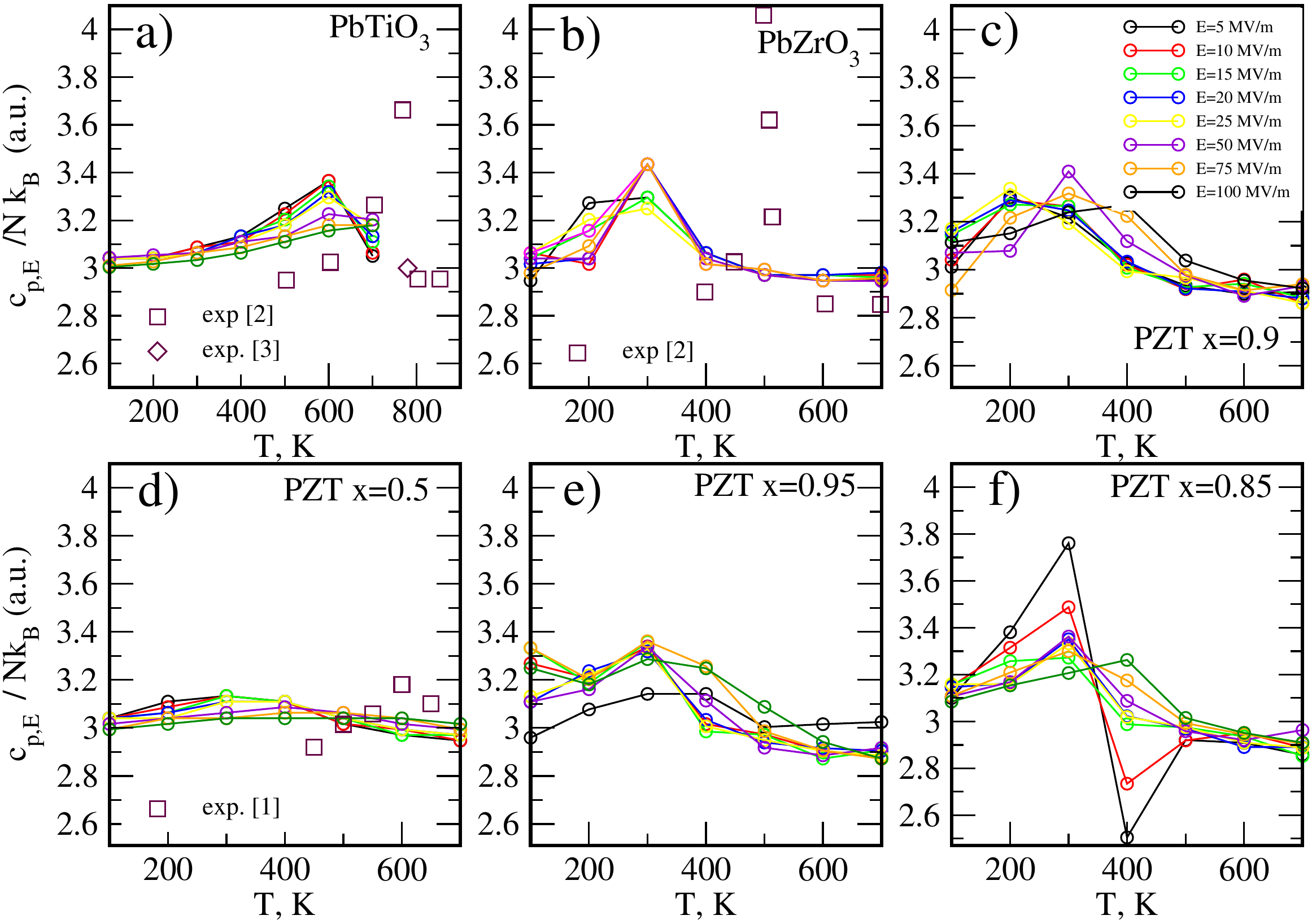}
    \caption{ Specific heat capacity calculated (open circles) for  (a) \PTO, (b) \PZO, (c) \PZTten, (d)  \PZTfi, (e) \PZTfive, (f) \PZTfifteen ~and experimental values (open squares and open diamonds) ~\cite{Morimoto2003, Yoshida2009, Rosetti1999}. 
   }
    \label{fig:SI1}
 \end{figure}

%%%%%%%%%%%%%%%%%%%%%%%%%%%%%%%%%%% 
%   AFE ECE
%%%%%%%%%%%%%%%%%%%%%%%%%%%%%%%%%%%  
\subsection{Calculated polarisation and electrocaloric temperature change in AFE \PZTfive, and AFE \PZTfifteen} 

Polarisation in  AFE \PZTfive, and AFE \PZTfifteen~ composites shows non-monotonic behaviour with critical field of 25 and 50 MV/m that turn the materials into monodomain at 300 and 400 K, respectively.
Calculated electrocaloric change in temperature for these AFE materials shows complex behaviour that related to domain competition.

%%%%%%  FIG2 %%%%%%
 \begin{figure}[h]
  \centering
  \includegraphics[clip=true,width=0.45\textwidth]{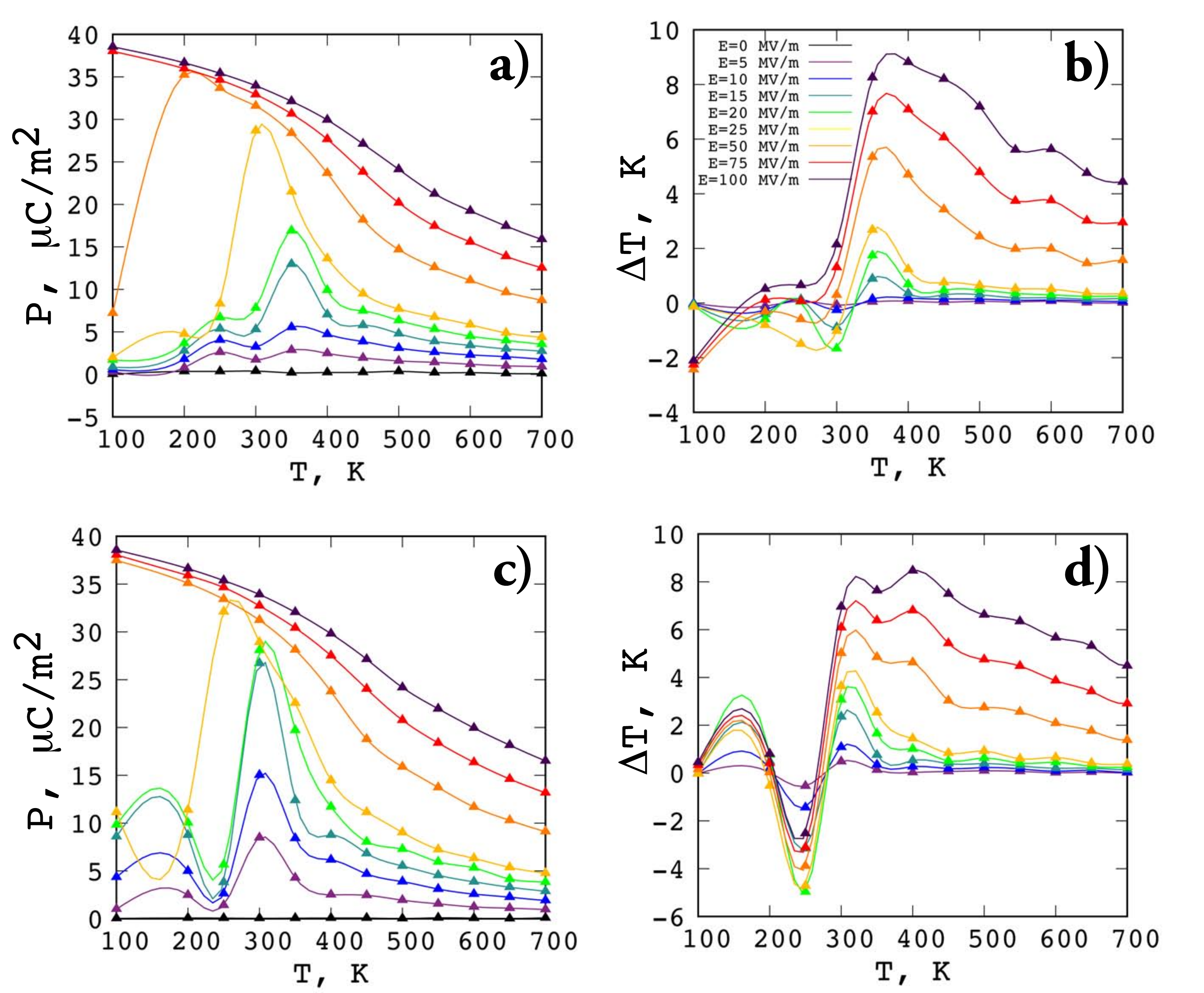}
    \caption{ Polarisation (a) \PZTfive, (c) \PZTfifteen~ and electrocaloric  change in temperature in (b)  \PZTfive, (d) \PZTfifteen~ for different temperatures and applied electric fields.
      }
    \label{fig:SI2}
 \end{figure}

%%%%%%%%%%%%%%%%%%%%%%%%%%%%%%%%%%% 
%   AFE and FE ECE
%%%%%%%%%%%%%%%%%%%%%%%%%%%%%%%%%%%  
\subsection{Calculated electrocaloric effect in  FE PbTiO$_3$, FE \PZTth, and AFE \PZTtw} 

Calculated electrocaloric change in temperature for FE \PZTth~  shows positive values similar to the behaviour of standard ferroelectric compounds (Fig. S3a). For AFE PZT compositions, $x$=0.8 $\Delta T$ demonstrates negative-to-positive crossover (Fig. S3b).

%%%%%%  FIG3 %%%%%%
 \begin{figure}[h]
  \centering
  \includegraphics[clip=true,width=0.25\textwidth]{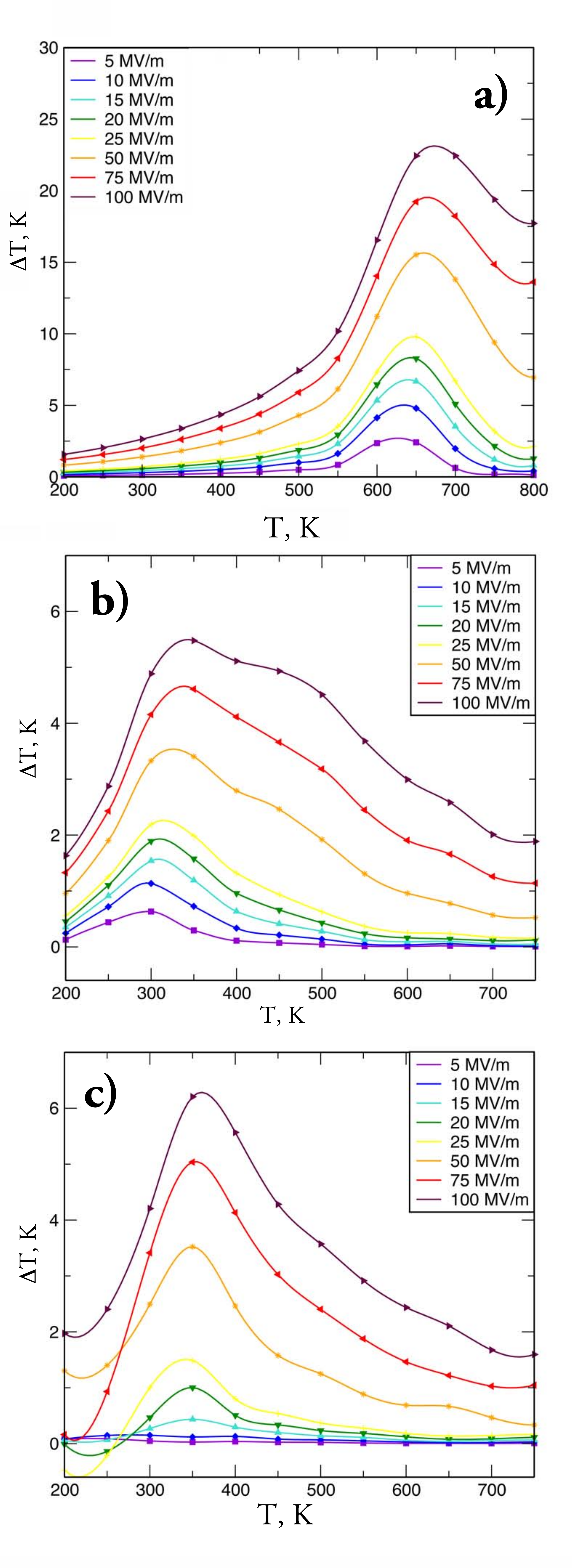}
    \caption{ The  electrocaloric  change in temperature in (a) PbTiO$_3$, (b) FE  \PZTth, and  (c) AFE \PZTtw~for different applied electric fields.
      }
    \label{fig:SI3}
 \end{figure}

%%%%%%%%%%%%%%%%%%%%%%%%%%%%%%%%%%% 
%   dP/dT and entropy change
%%%%%%%%%%%%%%%%%%%%%%%%%%%%%%%%%%%  
\subsection{Isothermal change of entropy in AFE and FE \PZT} 

The isothermal change of entropy,  $\Delta S$:
\begin{equation}%\label{eq:S1}
\Delta S=- \int \limits_{0}^{E}  \left(\frac{\partial P}{\partial T}\right)_E  \mathrm{d}E,          		    
\end{equation}
where $P$ is polarisation vector, $T$ is temperature, $E$ is electric field.
The change of polarisation with temperature and  isothermal change of entropy calculated  for AFE PZO,  \PZTfive, \PZTten, \PZTfifteen~ and,  for comparison, for FE \PZTfi~ compounds are shown in  Fig. S4. 
The FE system (Fig. S4 e and j) is characterised by small change of polarisation and the positive change of entropy leading to positive values of EC cage of temperature.
Meanwhile, AFE systems (Fig. S4 a-d and f-i) develops a relatively large change of $P$ with non-zero applied fields. Negative $\Delta S$ is related to positive values of  $\left(\frac{\partial P}{\partial T}\right)_E$, i.e. the increase of $P$ with temperature. 

%%%%%%  FIG4 %%%%%%
 \begin{figure}[h]
  \centering
  \includegraphics[clip=true,width=0.4\textwidth]{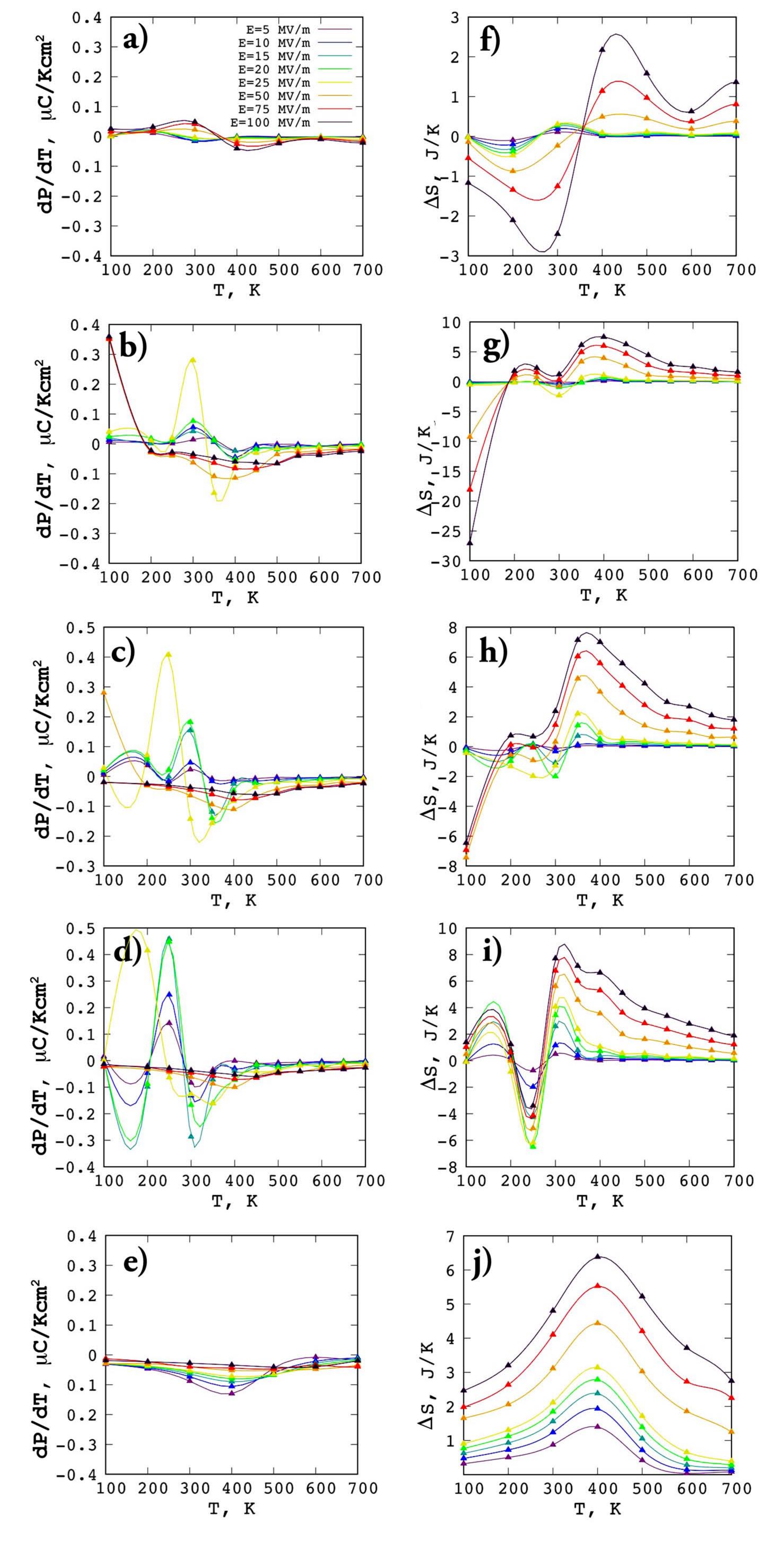}
    \caption{ The change of polarisation with temperature (a), (b), (c), (d), (e)  
                                  and  isothermal  change in entropy (f), (g), (h), (i),  (j)
                                  in \PZO, \PZTfive, \PZTten, \PZTfifteen, \PZTfi ~respectively,  for different applied electric fields.
      }
    \label{fig:SI4}
 \end{figure}

 \bibliography{ECE-revision.bib}